\documentclass{article}
\usepackage{amsmath}
\usepackage{amsfonts}
\usepackage{epsfig}
\usepackage{rotating}
\usepackage{graphicx}
\usepackage{amssymb}
\usepackage{eucal}
\usepackage{lscape}
\usepackage{latexsym,enumerate}
\usepackage{caption}
\usepackage{setspace}
\usepackage[margin=1.0in]{geometry} %uncomment

\usepackage{fancyhdr}
\usepackage{pdfpages}
\usepackage[super]{natbib}
\providecommand{\keywords}[1]{\textbf{\textit{Key words---}} #1}
\usepackage{xcolor}
\definecolor{green}{rgb}{0,0.75,0.25}

\newcommand{\cip}{\mbox{$\perp\!\!\!\perp$}}

\begin{document}

\title{Causal organic indirect and direct effects: closer to Baron and Kenny, with a product method for binary mediators}

%Running title: Causal organic indirect and direct effects

\author{Judith J.~Lok$^1$ and Ronald J.~Bosch$^2$\\
$^1$Department of Mathematics and Statistics,\\
Boston University,\\
jjlok@bu.edu\\
$^2$Center for Biostatistics in AIDS Research,\\
Harvard TH Chan School of Public Health\\
To appear in \emph{Epidemiology} 2021}

\maketitle

\begin{abstract} Mediation analysis, which started with Baron and Kenny~(1986), is used extensively by applied researchers. Indirect and direct effects are the part of a treatment effect that is mediated by a covariate and the part that is not. Subsequent work on natural indirect and direct effects provides a formal causal interpretation, based on cross-worlds counterfactuals: outcomes under treatment with the mediator set to its value without treatment. Organic indirect and direct effects (Lok 2016) avoid cross-worlds counterfactuals, using so-called organic interventions on the mediator while keeping the initial treatment fixed at treatment. Organic indirect and direct effects apply also to settings where the mediator cannot be set. In linear models where the outcome model does not have treatment-mediator interaction, both organic and natural indirect and direct effects lead to the same estimators as in Baron and Kenny (1986). Here, we generalize organic interventions on the mediator to include interventions combined with the initial treatment fixed at no treatment. We show that the product method holds in linear models for organic indirect and direct effects relative to no treatment even if there is treatment-mediator interaction. Moreover, we find a product method for binary mediators. Furthermore, we argue that the organic indirect effect relative to no treatment is very relevant for drug development. We illustrate the benefits of our approach by estimating the organic indirect effect of curative HIV-treatments mediated by two HIV-persistence measures, using ART-interruption data \emph{without} curative HIV-treatments combined with an estimated/hypothesized effect of the curative HIV-treatments on these mediators.
\end{abstract}


\section{Introduction: indirect and direct effects}\label{Intro}

Indirect and direct effects decompose the effect of a treatment $A$ on
an outcome $Y$ into 1.\ a part that is mediated through covariate $M$, the \emph{indirect effect}, and 2.\ a part that is not, the \emph{direct effect}. For example, treatment $A$ could be a blood pressure lowering medication, and the outcome $Y$ whether a person had a heart attack. The mediation questions: How much of the effect of treatment $A$ on heart attacks $Y$ is mediated by its effect on blood pressure $M$ ($A\rightarrow M\rightarrow Y$), and how much (if any) by other pathways? Another example: new HIV-curative therapies are being developed to reduce HIV-reservoirs while patients are on antiretroviral therapy (ART). The clinical effects of such therapies are assessed by ART withdrawal in the study participants, to identify off-ART viral control. The mediation question: What is the effect of an HIV-curative therapy $A$ on off-ART viral control $Y$, mediated by the effect of $A$ on the HIV-reservoir $M$, as measured by on-ART single-copy plasma HIV-RNA?

Mediation analysis\cite{BaronKenny} is especially important in the health sciences, like epidemiology and psychology. Knowing the assumptions under which these analyses are valid/causal is paramount. Several approaches to causal mediation analysis have been proposed.$^{2-12}$ \nocite{RobGreenmed,Pearlmed,Tyler,VanderWeelebook,imai2010identification,Ericmedsurv,Vanessa2,Geneletti,JamieThomas,didelez2019defining,Naimi} Important causal contributions are interpretation, extension of mediation analysis to various outcome types, and the inclusion of pre-treatment common causes of mediator $M$ and outcome $Y$, even for randomized treatment. Nguyen et al.\ (2019)\cite{nguyen2019clarifying} provide an overview of causal mediation analysis.

One disadvantage of most current causal interpretations of mediation analysis$^{2-7}$\nocite{RobGreenmed,Pearlmed,Tyler, VanderWeelebook,imai2010identification,Ericmedsurv} is the reliance on quantities such as: the outcome under treatment but with the mediator set to its value without treatment. These so-called cross-worlds counterfactuals rely on two simultaneous but different situations/worlds, treatment and no treatment, which never occur concurrently. The identifying assumptions are also cross-worlds, have been disputed,\cite{JamieThomas,didelez2019defining,robins2020interventionist} and lead some researchers to entirely forego causal mediation analysis. Randomized effects\cite{Vanessa2,vanderweele2014effect,vansteelandt2017interventional} avoid cross-worlds quantities, but still require setting the mediator after a random draw. Separable effects\cite{JamieThomas,didelez2019defining,robins2020interventionist} separate a treatment's effect on the mediator from its direct effect on the outcome, but still need all $Y^{(a,m)}$: the outcomes under treatment $a$ with the mediator set to $m$. Geneletti~(2007)\cite{Geneletti} considers stochastic interventions on the mediator, but without conditioning on common causes $C$ of mediator and outcome, unless effects conditional on $C$ or manipulating $C$ are of interest.

Organic indirect and direct effects\cite{medJL} provide causal mediation analyses that do not require setting the mediator, or cross-worlds quantities/assumptions. This article generalizes these organic indirect and direct effects. This generalization brings organic indirect and direct effects closer to the original approach to mediation analysis by Baron and Kenny.\cite{BaronKenny} It also leads to a product method for binary mediators. Furthermore, by combining organic interventions on the mediator with $a=0$, indirect effects can be estimated in data-scarce settings (Sections~\ref{promising} and~\ref{HIV}).

This article has four purposes: to bring organic indirect and direct effects to the attention of epidemiologists and clinical trialists, to bring causal mediation analysis closer to the original mediation approach, to introduce a product method for binary mediators, and to illustrate causal mediation analysis for selecting promising treatments for randomized trials.

\section{Setting and notation}

We start with randomized treatment $A$. Denote the pre-treatment common causes of mediator $M$ and outcome $Y$ by $C$. As is usual, see, e.g., VanderWeele~(2015)\cite{VanderWeelebook} page~464, assume there are no post-treatment common causes of $M$ and $Y$. This can be relaxed.\cite{medJLp} Throughout, superscript $^{(0)}$ indicates without treatment and superscript $^{(1)}$ indicates under treatment. Figure~1 illustrates the mediation set-up. Notice the absence of an arrow from $C$ into $A$ since treatment is randomized.

\includegraphics[scale=0.8,angle=0]{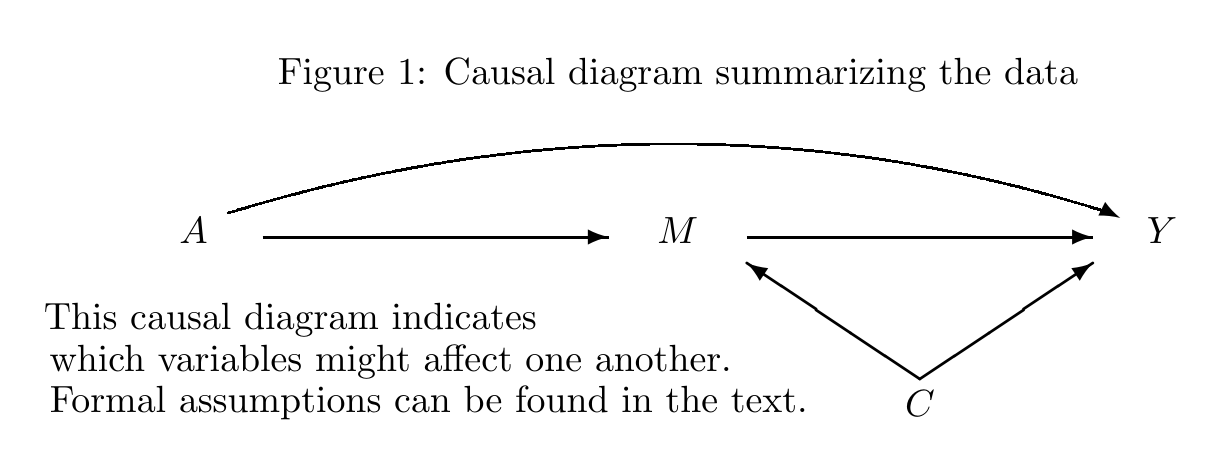}

\section{Definition of causal indirect and direct effects}\label{causal}

\subsection{Natural indirect and direct effects}

Historically$^{2-7}$\nocite{RobGreenmed,Pearlmed,Tyler,VanderWeelebook,imai2010identification,Ericmedsurv}, causal indirect and direct effects have been defined using the following counterfactuals: the outcomes under treatment had the mediator been set to the mediator level without treatment: $Y^{(1,M^{(0)})}$. There are two issues with these counterfactual outcomes $Y^{(1,M^{(0)})}$. First,
how to set the mediator is usually left unanswered, so the outcomes $Y^{(1,m)}$ under treatment with the mediator set to $m$ are undefined in many practical situations. When the mediator is another medication,\cite{Pearlmed,robins2020interventionist} setting the mediator is interpretable. When the mediator is a patient characteristic/covariate, as in many applications including our HIV-application, setting the mediator is harder to interpret. An illustrative example: ``There are many competing ways to assign (hypothetically) a body mass index of $25$ kg/m$^2$ to an individual, and each of them may have a different causal effect on the outcome''.\cite{ColeF}
Moreover, even if we could set the mediator, under treatment $M^{(0)}$ is not observed. Thus, it remains unclear how under treatment the mediator could be set to $M^{(0)}$.

Based on outcomes $Y^{(1,M^{(0)})}$, most current causal approaches to mediation analysis focus on the
natural indirect effect,
\begin{equation*}
E\left(Y^{(1)}-Y^{(1,M^{(0)})}\right),
\end{equation*}
mediated through $M$: it compares $M^{(1)}$ with $M^{(0)}$; and the natural direct effect,
\begin{equation*}
E\left(Y^{(1,M^{(0)})}-Y^{(0)}\right),
\end{equation*}
not mediated: $M$ is always $M^{(0)}$.

To estimate natural indirect and direct effects, since treatment $A$ is randomized, estimating $E\bigl(Y^{(1)}\bigr)$ and $E\bigl(Y^{(0)}\bigr)$ is standard. We focus on $E\bigl(Y^{(1,M^{(0)})}\bigr)$. Under strong conditions, with $C$ all pre-treatment common causes of mediator $M$ and outcome $Y$, the {\bf Mediation Formula}\cite{Pearlmed} holds:
\begin{equation}\label{MedFormNat}
E\left(Y^{(1,M^{(0)})}\right)=\int_{(m,c)}E\left[Y|M=m,C=c,A=1\right]f_{M|C=c,A=0}(m)f_C(c)dm\,dc.
\end{equation}
(\ref{MedFormNat}) was proven under cross-worlds (see Introduction) assumptions. In addition, most causal mediation approaches$^{2-12}$\nocite{RobGreenmed,Pearlmed,Tyler,VanderWeelebook,imai2010identification,Ericmedsurv,Vanessa2,Geneletti,JamieThomas,didelez2019defining,Naimi} need many counterfactual outcomes: not only
$Y^{(1,M^{(0)})}$, also all $Y^{(a,m)}$: the outcomes under treatment $a$ with the mediator set to $m$.

Under linear models and outcome models without exposure-mediator interaction, the resulting estimators for the indirect and direct effect are the same as in the original mediation approach,\cite{BaronKenny} and thus add a causal interpretation. Natural and randomized indirect and direct effects include all outcome types, with causal interpretations.

\subsection{Organic indirect and direct effects: an intervention-based approach avoiding cross-worlds counterfactuals/assumptions}\label{Organic}

We generalize the approach and main results from Lok (2016).\cite{medJL} $I$ is an intervention on the mediator that does not affect $C$ (pre-treatment common causes of mediator $M$ and outcome $Y$). $M^{(a,I=1)}$ and $Y^{(a,I=1)}$ denote the mediator and the outcome under treatment $A=a$ \emph{and} under intervention $I$ on the mediator. For observables without intervention on the mediator ($I=0$), we omit superscript $^{(I=0)}$.\medskip

\noindent {\bf Definition: \emph{(Organic intervention).}}
$I$ is an \textbf{\emph{organic intervention}} relative to $a=0$ and $C$
if
\begin{equation}\label{defint}
M^{(0,I=1)}\sim M^{(1)}\;\;\;\mid\;\;\;C=c
\end{equation}
and
\begin{equation}\label{ident}
Y^{(0,I=1)}\mid M^{(0,I=1)}=m,C=c \;\;\;\sim\;\;\;Y^{(0)}\mid M^{(0)}=m,C=c,
\end{equation}
where $\sim$ indicates having the same conditional distribution; that is, (\ref{ident}) states that the distribution of $Y^{(0,I=1)}$ given $M^{(0,I=1)}=m,C=c$ is the same as the distribution of $Y^{(0)}$ given $M^{(0)}=m,C=c$. $I$ is an organic intervention relative to $a=1$ and $C$ if (\ref{defint}) and (\ref{ident}) hold with the roles of $a=0$ and $a=1$ reversed.\\

\noindent (\ref{defint}) states that $I$ changes the distribution of the mediator to that under treatment, given $C$.
(\ref{ident}) states that $I$ ``has no direct effect
on the outcome'': without treatment, whether ${\rm mediator}=m$ came about due to $I$ or without $I$ does not affect a person's prognosis. 
Or, how the mediator came about is irrelevant. (\ref{ident}) can be relaxed to
\begin{equation*}
E\bigl[Y^{(0,I=1)}\mid M^{(0,I=1)}=m,C=c\bigr]=E\bigl[Y^{(0)}\mid M^{(0)}=m,C=c\bigr]:
\end{equation*}
whether ${\rm mediator}=m$ came about due to $I$ or without $I$ does not affect a person's mean prognosis. A similar relaxation holds for effects relative to $a=1$.\cite{medJL}

For example,\cite{medJL} $A=1$ could be a blood pressure lowering medication, $M$ a person's blood pressure, and $Y$ the subsequent occurrence of a heart attack. The mediation questions: what is the effect of $A=1$ mediated by its effect on blood pressure, and does $A=1$ also have a direct effect on heart attacks? It could be that $A=1$ lowers blood pressure by 10, on average for each $C$, without changing the shape of the blood pressure distribution. $I$ should then be an intervention, in the untreated, that also lowers blood pressure by 10, on average for each $C$, without changing the shape of the blood pressure distribution. Then,
$M^{(0,I=1)}\sim M^{(1)}\mid C$. $I$ could be salt reduction in a dosage dependent on $C$. Since the effect of salt on heart attacks is believed to be through its effect on blood pressure,\cite{CDCbloodpressure,Grillo2019}
one can posit that (\ref{ident}) holds
for salt reduction intervention $I$.  
Salt reduction $I$ keeps the effect of $A=1$ through blood pressure $M$, but does not have the direct effect of $A=1$.

For (\ref{ident}) to hold for any intervention $I$, one will often need the pre-treatment common causes of mediator $M$ and outcome $Y$ in $C$:\cite{medJL} without $C$, the statement ``the mediator under no treatment equals $m$'' likely implies a different prognosis under intervention $I$ ($M^{(0,I=1)}=m$) versus without intervention $I$ ($M^{(0)}=m$), because $M^{(0)}$ and $M^{(0,I=1)}$ are related differently to $C$, so convey different information about $C$, which may well predict the outcome $Y$.\medskip

\noindent {\bf Definition: \emph{(Organic indirect and direct effects relative to $a=0$ and $C$):}}
\begin{equation*}
E\bigl(Y^{(0,I=1)}\bigr)-E\bigl(Y^{(0)}\bigr)
\end{equation*}
is the \textbf{\emph{organic indirect effect}} of treatment $A$, relative to $a=0$ and $C$, based on $I$. The treatment is $0$ for both $Y^{(0)}$ and $Y^{(0,I=1)}$, so this effect is mediated.
\begin{equation*}
E\bigl(Y^{(1)}\bigr) -E\bigl(Y^{(0,I=1)}\bigr)
\end{equation*}
is the \textbf{\emph{organic direct effect}} of treatment $A$, relative to $a=0$ and $C$, based on $I$. The mediator has the same distribution for $Y^{(1)}$ and $Y^{(0,I=1)}$, so this effect is not mediated.

\noindent {\bf \emph{Organic indirect and direct effects relative to $a=1$ and $C$}} are defined similarly, with the roles of $a=0$ and $a=1$ reversed.\medskip

\noindent Combining organic interventions $I$ with no treatment provides the effect of an intervention $I$ that affects the mediator $M$ the same way as $A=1$, but has no direct effect on the outcome $Y$.
Such effect can be highly relevant for drug development; see Sections~\ref{promising} and~\ref{HIV}.

The Mediation Formula holds for organic interventions\cite{medJL} (eAppendix~\ref{Amedform}):

\noindent {\bf Theorem}: \textbf{\emph{Organic indirect and direct effects: the Mediation Formula.}} Under randomized treatment, for an intervention $I$ that is organic relative to $a$ and $C$:
\begin{equation}E\bigl(Y^{(a,I=1)}\bigr)
=\int_{(m,c)}E\left[Y|M=m,C=c,A=a\right]f_{M|C=c,A=1-a}(m)f_C(c)dm\,dc.
\label{medform}
\end{equation}

\noindent The Mediation Formula implies that for given $C$, organic indirect and direct effects don't depend on the intervention $I$, as long as $I$ is organic. The resulting expression is in terms of observable quantities only. All ingredients can be estimated using standard methods (eAppendix~\ref{Aexa}); the empirical distribution function of $C$ estimates $f_C$, so we estimate $E\left(Y^{(a,I=1)}\right)$ by the average of the integrand over the $C_i$.

Furthermore, if $C$ has all pre-treatment common causes of mediator $M$ and outcome $Y$, as defined in eAppendix~\ref{Aunique}, organic indirect and direct effects are unique: which set of all pre-treatment common causes $C$ is chosen does not affect organic indirect and direct effects\cite{medJL} (eAppendix~\ref{Aunique}). Thus, we can define:\medskip

\noindent {\bf Definition:} Whenever $C$ includes all common causes of mediator $M$ and outcome $Y^{(a)}$, we call organic indirect and direct effects relative to $a$ and $C$ {\bf \emph{the} organic indirect and direct effects relative to $a$}.\medskip

Organic indirect and direct effects relative to $a=1$ generalize natural indirect and direct effects: in settings where an intervention exists that sets the mediator under $a=1$ to the value it would have taken under $a=0$, that is, $M^{(1,I=1)}=M^{(0)}$ exists, the usual cross-worlds assumptions, see e.g.,~pages~463-464 in VanderWeele 2015,\cite{VanderWeelebook} imply that this $I$ is organic relative to $a=1$.\cite{medJL} This approach also provides a proof of the mediation formula for natural indirect and direct effects under conditions that are somewhat weaker than usual (eAppendix~\ref{Aweaker}).

In linear models without treatment-mediator interaction in the outcome model, the product method\cite{BaronKenny} holds for all these causal estimands: natural, randomized, and organic effects.$^{3-8;10;17}$\nocite{Pearlmed,Tyler,VanderWeelebook,imai2010identification,Ericmedsurv,Vanessa2,JamieThomas,medJL} When the outcome model has treatment-mediator interaction, the product method holds for organic indirect effects relative to $a=0$ (eAppendix~\ref{Aproduct}), and for natural and randomized indirect effects after reversing the roles of $a=0$ and $a=1$, leading to pure indirect effects.\cite{RobGreenmed}

\section{A product method for binary mediators}\label{ProdSec}

For binary mediators, the organic indirect effect relative to $a=0$, and the pure indirect effect, equal
\begin{eqnarray}\lefteqn{\int_{c}\left(E\left[Y|M=1,C=c,A=0\right]-E\left[Y|M=0,C=c,A=0\right]\right)}\nonumber\\
&&\left(P(M=1|C=c,A=1)-P(M=1|C=c,A=0)\right)f_C(c)\,dc\label{productBin}
\end{eqnarray}
(eAppendix~\ref{Aexa}): a product method for binary mediators.

If there are no common causes $C$, or mediation conditional on $C$ is of interest as in Chapters~2-7 of~VanderWeele~(2015),\cite{VanderWeelebook} (\ref{productBin}) is a true product. Otherwise, indirect effects are indirect effects conditional on $C$ (a true product), averaged over $C$. While averaging over $C$ may result in a complicated integral, for estimation one can use the empirical average over all observed $C_i$.
The product method holds for binary mediators regardless of outcome type; for binary outcomes, it holds on the risk difference scale.

Intuitively, the indirect effect relative to $a=0$ is the product of the treatment effect on the mediator times the effect of this mediator change on the outcome under $A=0$, averaged over $C$. If $A$ increases the probability of $M=1$, the indirect effect relative to $a=0$ is the increased probability of $M=1$ times the effect that changing $M=0$ into $M=1$ has on the outcome under $A=0$, averaged over $C$. This is similar to the original product method.

A similar product method holds for organic indirect effects relative to $a=1$ and natural indirect effects; in the conditional expectation of $Y$, $A=1$ replaces $A=0$ (eAppendix~\ref{Aexa}). Note that by (\ref{productBin}) there is a maximum indirect effect that can be mediated through a binary mediator.

\section{Selecting new treatments with promising indirect effects for clinical trials}\label{promising}

From the Mediation Formula (\ref{medform}), the organic indirect effect relative to $a=0$ equals
\begin{equation*}\int_{(m,c)}E\left[Y|M=m,C=c,A=0\right]\left(f_{M|C=c,A=1}(m)-f_{M|C=c,A=0}(m)\right)f_C(c)dm\,dc
\end{equation*}
\begin{equation}\label{indir} \text{or   }\int_{(m,c)}E\left[Y|M=m,C=c,A=0\right]f_{M|C=c,A=1}(m)f_C(c)dm\,dc-E\bigl(Y^{(0)}\bigr).
\end{equation}
Thus, the organic indirect effect relative to $a=0$ can be estimated with 1.\ the distribution of the mediator $M$ under treatment, $A=1$, and without treatment, $A=0$, and 2.\ the expected outcome $Y$ given the mediator $M$ and pre-treatment common causes $C$ without treatment, $A=0$; expected outcomes \emph{only} under $A=0$.

The same holds for natural and randomized indirect effects after reversing the roles of $a=0$ and $a=1$, that is, for pure indirect effects. This observation allows estimating \emph{indirect} effects relative to $a=0$ of potential new treatments aiming to affect a mediator $M$,\cite{Frank} without observing treated outcomes (Section~\ref{HIV}). eAppendix~\ref{Aexa} provides estimators of (\ref{indir}) under various model specifications.

Similarly, but less practically relevant, organic indirect effects relative to $a=1$ and natural indirect effects could be estimated without observing untreated outcomes.

\section{Selecting HIV-curative treatments with promising indirect effects for clinical trials}\label{HIV}

ART, the standard of care for HIV, has rendered HIV a chronic disease. Substantial research now focuses on HIV-eradication, or functional cure aimed at long-term ART-free HIV-remission.\cite{margolis2016latency} Trials interrupt ART in HIV-infected study participants, to investigate the effect of curative treatments and on-ART biomarkers on the time to viral rebound after stopping ART. Viral rebound is the time the HIV-viral load in the participants' blood rises above a pre-specified level.\cite{Li2015} 

ART-interruption trials have to be carried out with extreme care, since ART-interruption carries significant risks.$^{24-27}$\nocite{Li2015,SMART,Li2014,julg2019recommendations} Because there are many potential treatments in pre-clinical stages, it is advantageous to carry out ART-interruption trials for the most promising therapies.\cite{Ghosn}

Many new curative HIV-therapies $A$ are designed to affect on-ART HIV-persistence, let's tentatively call its measurement $M$, which has an effect on the time to viral rebound, $Y$ (Figure~2). For such therapies, equation~(\ref{indir}) leads to an estimate of their organic \emph{indirect} effect relative to $a=0$ mediated by $M$, with $M$ single-copy plasma HIV-RNA or cell-associated HIV-RNA. Pre-ART HIV-viral load measures the magnitude of viral replication before ART initiation, which is predictive of virologic outcomes after stopping ART, $Y$,\cite{Treasure2016} and of on-ART HIV-persistence measures, $M$.\cite{Riddler2015,Ghandi2017} Without information on pre-ART viral load, nadir CD4 count is our surrogate $C$: a potential pre-treatment common cause of mediator and outcome.\cite{Li2015}

\hspace*{-0.7cm}\includegraphics[scale=0.88,angle=0]{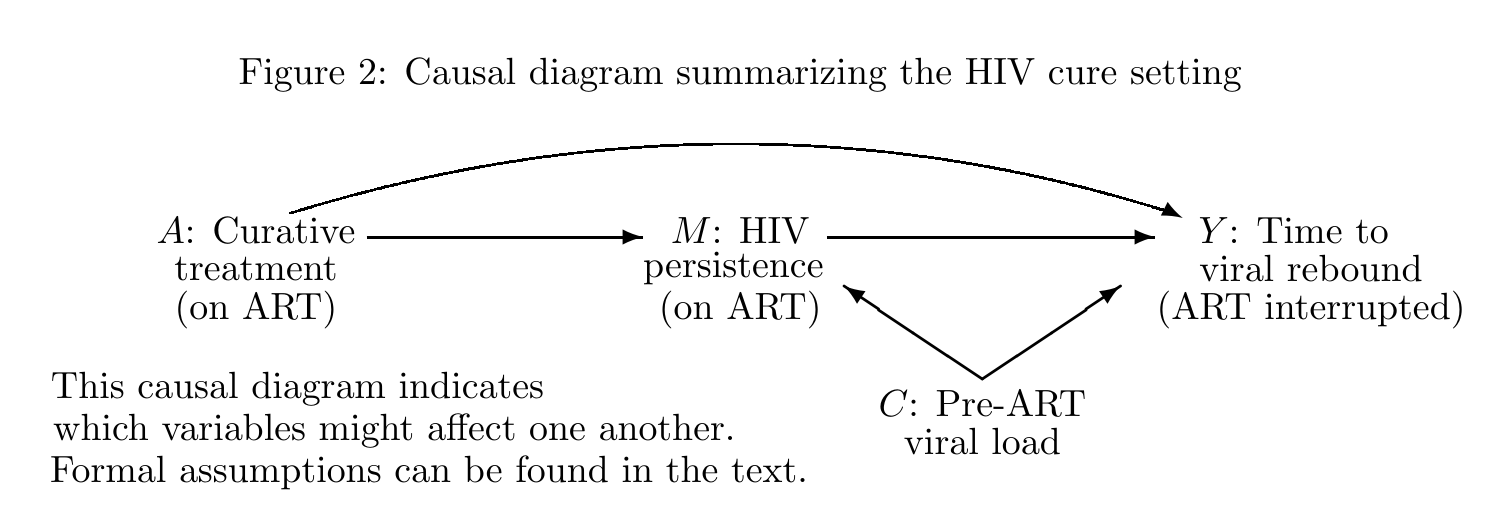}

We use data from completed ACTG trials:\cite{Li2015} $124$ HIV-infected study participants without curative treatments and with ART-interruptions, $(Y_i,M_i,C_i)$, $i=1,\ldots,n$ (assumed independent and identically distributed). Combining these existing data, all with $A_i=0$, with hypothesized effects of potential new curative HIV-treatments $A$ on biomarkers $M$, we estimate the indirect effect relative to $a=0$ of new treatments $A$: the effect of organic interventions with the same effect on the mediator as such new treatments $A=1$, and no effect on $Y$ through other pathways.

We categorize the outcome viral rebound to suppressed ($Y=1$)/not suppressed ($Y=0$) through week 4 of ART-interruption.  Let $\bar{M}=1$ if the HIV-persistence measure is below the assay limit, and $\bar{M}=0$ if it is above. First, we estimate the indirect effect of a treatment $A$ that increases by 3-fold the odds that the HIV-persistence measure is below the assay limit, given $C$. The indirect effect mediated by the binary on-ART HIV-persistence measure $\bar{M}$ equals (equation~(\ref{indir})):
\begin{eqnarray*}\lefteqn{\int_{c}P\left(Y=1|\bar{M}=1,C=c,A=0\right)P(\bar{M}^{(1)}=1|C=c)f_C(c)dc}\\
&&+\int_{c}P\left(Y=1|\bar{M}=0,C=c,A=0\right)P(\bar{M}^{(1)}=0|C=c)f_C(c)dc-P\bigl(Y^{(0)}=1\bigr).
\end{eqnarray*}
$P(\bar{M}^{(1)}=m|C=c)$ and $P\left(Y=1|\bar{M},C,A=0\right)$ can be estimated from the data,\cite{Li2015} $(Y_i,\bar{M}_i,C_i)$, $i=1,\ldots,n$, all with $A_i=0$, under the hypothesized effect of treatment on the mediator. We fit a logistic regression model
\begin{equation}
\text{logit } P(\bar{M}_i=1|C_i=c,A_i=0)=\text{logit }p_{0,\eta}(c)=\eta_0+\eta_1c.
\end{equation}
If treatment increases this odds 3-fold, 
\begin{equation*}
\hat{P}_{\hat{\eta}}(\bar{M}^{(1)}=1|C=c)=p_{1,\hat{\eta}}(c)=3p_{0,\hat{\eta}}(c)/(1+(3-1)p_{0,\hat{\eta}}(c))
\end{equation*}
(eAppendix~\ref{Odds}).
We also fit a logistic regression model
\begin{equation}\label{betahatbar}\text{logit }P\left(Y_i=1|\bar{M}_i=m,C_i=c,A_i=0\right)=\text{logit }p_{\alpha}(m,c)=\alpha_0+\alpha_1m+\alpha_2c.
\end{equation}
The indirect effect is estimated by
\begin{equation*}\frac{1}{n}\sum_{i=1}^n\sum_{m=0}^1p_{\hat{\alpha}}\left(Y=1|\bar{M}=m,C=c_i,A=0\right)P_{\hat{\eta}}(\bar{M}^{(1)}=m|C=c_i)-\frac{1}{n}\sum_{i=1}^ny_i,
\end{equation*}
that is,
\begin{equation*}\frac{1}{n}\sum_{i=1}^n\left(p_{\hat{\alpha}}(1,c_i)\frac{3p_{0,\hat{\eta}}(c_i)}{1+(3-1)p_{0,\hat{\eta}}(c_i)}+p_{\hat{\alpha}}(0,c_i) \frac{1-p_{0,\hat{\eta}}(c_i)}{1+(3-1)p_{0,\hat{\eta}}(c_i)}
-y_i\right).
\end{equation*}
Table~\ref{tab:bin} summarizes the results.

Next, consider the original HIV-persistence measure $M$, usually measured on the log$_{10}$ scale. Mathematical models predicted that a 10-fold HIV-reservoir reduction increases the time to viral rebound by only a few weeks to a month.\cite{Hill2014} This has not been confirmed in clinical studies, which have seen smaller reservoir reductions.\cite{Hill2014} We estimate the organic indirect effect relative to $a=0$ of a new curative treatment $A$ that leads to a 10-fold HIV-reservoir reduction. On the log$_{10}$ scale, such $A$ shifts the mediator distribution $M$ by one-log$_{10}$ given $C$:
\begin{equation}\label{shift}
M^{(1)}\sim M^{(0)}-1\,{\rm log}_{10}\mid C.
\end{equation}
Treatment $A$ shifts the \emph{distribution} of the mediator rather than the actual values: we don't assume rank preservation.\cite{Enc,MimAn} The distribution of $M^{(1)}$ is estimated by subtracting one-log$_{10}$ from the mediator under no treatment $M^{(0)}$.

The organic indirect effect relative to $a=0$ of a treatment $A$ that shifts the distribution of the mediator by one-log$_{10}$ given $C$ equals (equation~(\ref{indir}))
\begin{eqnarray}\label{indirHIV}\lefteqn{\int_{(m,c)}P\left(Y=1|M=m,C=c,A=0\right)f_{M|C=c,A=1}(m)f_C(c)dm\,dc-P\bigl(Y^{(0)}=1\bigr)}\nonumber\\
&=&\int_{(m,c)}P\left(Y=1|M=m,C=c,A=0\right)f_{M|C=c,A=0}(m+1\, log_{10})f_C(c)dm\,dc-P\bigl(Y^{(0)}=1\bigr)\nonumber\\
&=&\int_{(m,c)}P\left(Y=1|M=\tilde{m}(m),C=c,A=0\right)f_{M|C=c,A=0}(m)f_C(c)dm\,dc-P\bigl(Y^{(0)}=1\bigr),
\end{eqnarray}
with a change-of-variables argument (eAppendix~\ref{Aexa}), where $\tilde{m}(m)=m-1\,log_{10}$ if $m-1\,log_{10}$ is above the assay limit, and $\tilde{m}(m)$ is ``below the assay limit'' otherwise. We fit the logistic regression model
\begin{eqnarray}\label{betahat}\text{logit }P\left(Y_i=1|M_i=m, C_i=c, A_i=0\right)&=&\text{logit }p_{\beta}(m,c)\nonumber\\
&=&\left\{\begin{array}{ll}\beta_0+\beta_1m+\beta_2c&\text {if}\;m\;\text{above the assay limit}\\ 
\beta_3+\beta_2c&\text{if}\;m\;\text{below the assay limit.}
\end{array}\right. 
\end{eqnarray}
The organic indirect effect (\ref{indirHIV}) is estimated by
\begin{equation*}\frac{1}{n}\sum_{i=1}^n\left(p_{\hat{\beta}}(\tilde{m}(m_i),c_i) -y_i\right).
\end{equation*}
eAppendix~\ref{AppHIVResults} fits outcome and mediator models. Table~\ref{tab:cont} summarizes the results.

From Table~\ref{tab:cont}, a one-log$_{10}$ decrease in CA-HIV-RNA may lead to a more promising indirect effect on viral control than a one-log$_{10}$ decrease in SCA-HIV-RNA, and CA-HIV-RNA may thus be a better target for curative HIV-treatments. This may however be influenced by measurement error, present for both mediators, and/or due to a larger fraction of SCA-HIV-RNAs below the assay limit without curative treatment (60/94 (64\%) versus 52/124 (42\%)).

If we have measurements of the on-ART biomarker $M^{(1)}$ and $C$ for HIV-infected individuals under a curative treatment, say $(m^{(1)}_{j},c^{(1)}_{j})_{j=1}^{J}$ in a population similar to our study population, but not necessarily ART-interruption data for the new treatment, the organic indirect effect is estimated by
\begin{equation}\frac{1}{J}\sum_{j=1}^Jp_{\hat{\beta}}(m^{(1)}_{j},c^{(1)}_{j})-\frac{1}{n}\sum_{i=1}^ny_i,\label{M1meas}
\end{equation}
with $p_\beta$ and $\hat{\beta}$ as before, estimated in patients with $A_i=0$.
eAppendix~\ref{Mmeasurements} describes settings with on-treatment measurements of $C$ and $M$ in a different population.

\section{Discussion}

For organic indirect and direct effects, one doesn't need to envision setting the mediator; one needs to envision affecting the \emph{distribution} of the mediator, which potential new treatments often aim to do. Our initial approach\cite{medJL} generalized natural$^{2-7}$\nocite{RobGreenmed,Pearlmed,Tyler,VanderWeelebook,imai2010identification,Ericmedsurv} and randomized\cite{Vanessa2,vanderweele2014effect,vansteelandt2017interventional} indirect and direct effects by combining organic interventions with treatment, $a=1$. When combining organic interventions on the mediator with $a=0$, the no treatment scenario, organic indirect effects relative to $a=0$ and pure indirect effects\cite{RobGreenmed, nguyen2019clarifying} only depend on the model for $E[Y|M,A,C]$ restricted to $a=0$, where interactions between treatment and mediator are irrelevant. A product method even holds for binary mediators (Section~\ref{ProdSec}).

Intervention-based reasoning helps determine whether combining interventions on the mediator with $a=0$ or with $a=1$ is most practically relevant. For the desired effect of a treatment through the mediator, as in our HIV-cure example (Section~\ref{HIV}) or the effect of AZT on HIV-transmission\cite{medJL} (eAppendix~\ref{AHIVtrans}), combining interventions on the mediator with $a=0$ is interesting. If interest lies in preventing a treatment's side effects $M$, combining interventions on the mediator with $a=1$ is interesting. VanderWeele~(2015)\cite{VanderWeelebook} pages~193-196 describes related work on three-way decompositions.

Under the respective conditions usually made and if there are no post-treatment common causes of mediator and outcome, natural direct and indirect effects, their randomized counterparts,\cite{Vanessa2,vanderweele2014effect,vansteelandt2017interventional} and their organic counterparts relative to  $a=1$ all lead to the same numerical results in settings where these effects are all well-defined: the respective interventions are organic under those conditions\cite{medJL} (eAppendix~\ref{Aweaker}). This also proves that randomized indirect and direct effects don't depend on the choice of all common causes $C$ of mediator and outcome (eAppenxix~\ref{Aunique}). Organic indirect and direct effects relative to $a=0$ can differ from those relative to $a=1$, and similarly lead to the same numerical results as for natural and randomized indirect and direct effects when reversing also there the roles of $a=1$ and $a=0$.

Organic \emph{indirect} effects relative to $a=0$ can be estimated without outcome data under treatment. Section~\ref{HIV} illustrates this is useful for selecting potential new treatments with promising indirect effects for further evaluation in clinical trials. There are often many candidate treatments $A$ targeting a process which can be measured by a biomarker $M$. Data on clinical outcomes and mediators under no treatment may be available from observational studies, or from placebo arms of completed studies. Our proposed method is particularly useful in data-scarce settings, like the current evaluation of COVID-19 treatments. Combining the hypothesized treatment effect on biomarker $M$ (potentially informed by early-phase studies or pre-clinical animal studies) with outcomes under no or ineffective treatment, one could estimate the potential indirect effect mediated by $M$ relative to $a=0$ using (\ref{indir}), as in our HIV-application. Randomized trials could be reserved for treatments with the most promising indirect effects.

It may sound counterintuitive that we don't need outcomes under treatment, and the treatment effect on the mediator suffices to estimate organic indirect effects relative to $a=0$ and pure indirect effects. However, this only provides indirect effects, not total effects or the proportion mediated. For organic indirect effects relative to $a=0$, one can think of the mediator being affected and then the outcome taking its course as under $A=0$; for that, outcome data under treatment are not needed. If outcome data under treatment are available, for outcomes under $A=1$ to affect predictions from a full model $E[Y|C,A,M]$ restricted to $a=0$ (and thus indirect effects), $E[Y|C,A,M]$ has to include parameters that are common for $A=0$ and $A=1$. In fact, the same reasoning holds for pure indirect effects. In contrast, depending again on the outcome model specification, for the natural indirect effect, outcome data without treatment might not affect the estimated indirect effect.

A key lesson from the HIV-example is the importance of collecting pre-treatment common causes $C$ of mediator and outcome. Even in randomized trials, data on pre-treatment common causes $C$ is needed to estimate indirect and direct effects. The indirect effect is expected to not be large if $C$, not $M$, is the main predictor of $Y$. A sensitivity analysis for missing $C$ exists for linear structural equation models.\cite{imai2010identification}

In our HIV-example, the probability of virologic suppression was assumed not to depend on how far below the assay limit a mediator value lies. If new treatments also lower the mediator further below the assay limit, the indirect effect relative to $a=0$ of new treatments could be larger than we estimated. An interesting topic for future research is estimating the impact of an intervention that shifts the mediator, incorporating that mediators further below the limit could lead to longer times to viral rebound.

As usual in mediation analysis, when the mediator $M$ is just a marker for some underlying disease process, it cannot be expected that the distribution of the outcome given $M^{(0,I=1)}=m$ and $C$ is similar to the distribution of the outcome given $M^{(0)}=m$ and $C$, unless the intervention $I$ affects the marker because it affects the underlying disease process. A famous counter-example is the effect of yellow fingers on lung cancer. ``Yellow fingers'' would constitute an irrelevant mediator, unless used as a measure of smoking intensity. The meaning of mediation analysis is distorted when the interpretation of the potential mediator is different under treatment than without treatment. For example, in HIV-studies\cite{Bosch2013,insight2009interleukin} IL-2 increased CD4 counts, and in prior ART treatment trials, CD4 count increases  predicted better clinical outcomes. However, IL-2 didn't improve clinical outcomes in randomized trials. 
Stated in terms of interventions, when the CD4 count mediator came about due to IL-2, it had a different prognostic effect on the outcome than when it came about without IL-2: IL-2 is not an organic intervention. This is further illustrated by the fact that CD4 counts resulting from IL-2 can be much higher than typically seen in any HIV-infected patient, resulting in equation (\ref{ident}) conditioning on null events. Mechanism of action and subject-matter knowledge are important considerations for causal mediation analysis.

If treatment results in mediator values rarely seen without treatment, the organic indirect effect relative to $a=0$ is unidentified: some conditioning events in the Mediation Formula are null events, events that never happen, so the Mediation Formula is ambiguous. Any effects identified by the Mediation Formula, including natural, randomized,\cite{Vanessa2,vanderweele2014effect,vansteelandt2017interventional} and separable\cite{JamieThomas,didelez2019defining,robins2020interventionist} effects, suffer from this phenomenon. Extrapolation using parametric models can render results very model-dependent.\cite{ho2007matching}

Since the Mediation Formula holds for organic indirect and direct effects, many results for natural indirect and direct effects generalize to organic indirect and direct effects. For example, organic indirect and direct effects can be estimated when there is measurement error.\cite{Valeri} If there is measurement error in the mediator, as in our HIV-example, the indirect effect is typically underestimated.\cite{Valeri} Measurement error in the mediator combined with mediator assay lower limits is an interesting topic for future research. 

The proposed organic indirect and direct effects can also be estimated from observational data, if all confounders have been measured\cite{medJL} (eAppendix~\ref{Aobs}). There exist preliminary results on incorporating post-treatment common causes of mediator and outcome\cite{vanderweele2014effect} into our framework.\cite{medJLp}

\section*{Computer code and data}

\noindent SAS 9.4 (SAS Institute Inc., Cary, NC, USA) computer code for the analyses is available on the first author's website, http://www.bu.edu/math/judith-lok/. R-code for the point estimates, developed by Ariel Chernofsky, is available on https://github.com/a-chernofsky/lok-bosch-2020/. The details of the data collection, including single copy and cell-associated RNA methods, can be found in Li et al. (2016). The data used in Section~7 is available upon request from the first author.

\section*{Acknowledgements}

The authors are extremely grateful to the HIV-infected participants who volunteered for the ACTG ART interruption trials. The authors also thank Dr.~J.Z.~Li (Brigham and Women Hospital, Boston) for his support, for leading the ART interruption project and for generating the mediator data we analyzed. The authors are also grateful to Ariel Chernofsky (PhD student at the Department of Biostatistics, Boston University) for re-programming the point estimates for our HIV application, both for binary mediators and for mediator shifts, in R, in order to validate our SAS code. We thank the anonymous referees for \emph{Epidemiology} for encouraging us to define \emph{the} organic indirect and direct effect as any version where $C$ has all common causes, and for various other useful suggestions.

\section*{Sources of Funding}

\noindent The results reported herein correspond to Aim 1 of grant DMS 1810837 to Judith J.~Lok from the National Science Foundation. This work was also supported by NIH/NIAID grant UM1 AI068634 and, through generating the data, AI068636. %Ron's grant and the ACTG grant.
The content is solely the responsibility of the authors and does not necessarily represent the official views of the NSF or the NIH. 

\section*{Research Ethics and Informed Consent}

The authors have IRB approval at their respective institutions for this research. The participants of the studies analyzed in Section~7 gave informed consent and the studies were approved at the participating institutions.

\addcontentsline{toc}{chapter}{Bibliography}
\bibliographystyle{Vancouver} \bibliography{ref}

\begin{landscape}
\begin{table}
\caption{Organic indirect effects of curative HIV-treatments that increase the odds of HIV-reservoir measures below the assay limit.\\
The estimated probability of virologic suppression without treatment was $63/124$ or $51\%$ (week 4) and $17/122$ or $14\%$ (week 8). }
\begin{tabular}{lllllll}
Binary mediator & OR of below & Week$^{b,c}$ & Indirect effect$^d$ & 95\% CI$^e$ & RR$^f$ & 95\% CI$^e$\\
&assay limit$^a$\\ 
\hline\\
SCA HIV-RNA$^g$& 2 & 4 & 2.4\% & (-0.5\%,5.1\%) & 1.04 & (0.99,1.10)\\
SCA HIV-RNA$^g$& 3 & 4 & 3.5\% & (-0.7\%,7.5\%) & 1.06 & (0.99,1.14)\\
SCA HIV-RNA$^g$& $\infty$$^i$ & 4 & 6.3\% &(-1.3\%,14.1\%) & 1.12 & (0.98,1.27)\\
\\
CA HIV-RNA$^h$& 2 & 4 & 3.7\% & (0.86\%,6.4\%) & 1.07 & (1.02,1.13)\\
CA HIV-RNA$^h$& 3 & 4 & 5.7\% & (1.3\%,9.9\%) & 1.11 & (1.03,1.21)\\
CA HIV-RNA$^h$& 10 & 4 & 10.0\% & (2.3,17.5\%) & 1.20 & (1.05,1.36)\\
CA HIV-RNA$^h$& $\infty$$^i$ & 4 & 12.7\% & (3.0\%,22.7\%) & 1.25 & (1.06,1.47)\\
\\
SCA HIV-RNA$^g$& 2 & 8 & 2.2\% & (0.4\%,4.1\%) & 1.14 & (1.03,1.25)\\
SCA HIV-RNA$^g$& 3 & 8 & 3.2\% & (0.5\%,6.0\%) & 1.20 & (1.04,1.36)\\
SCA HIV-RNA$^g$& $\infty$$^i$ & 8 & 5.7\% &(0.9\%,11.1\%) & 1.35 & (1.07,1.68)\\
\\
CA HIV-RNA$^h$& 2 & 8 & 4.0\% & (1.7\%,6.3\%) & 1.29 & (1.15,1.43)\\
CA HIV-RNA$^h$& 3 & 8 & 6.4\% & (2.8\%,10.0\%) & 1.46 & (1.23,1.69)\\
CA HIV-RNA$^h$& 10 & 8 & 11.9\% & (5.1\%,18.8\%) & 1.86 & (1.43,2.36)\\
CA HIV-RNA$^h$& $\infty$$^i$ & 8 & 15.9\% & (6.9\%,25.6\%) & 2.14 & (1.56,2.92)\\
\\
\end{tabular}\\
\label{tab:bin}
\noindent$^a$ Odds Ratio (OR) the mediator is below the assay limit, given the common cause $C$, relative to no curative treatment $a=0$.\\
$^b$ For the week-4 analysis, analyzing viral rebound in the first 4 weeks after ART treatment interruption, $C$ is NNRTI-based (versus not).\\
$^c$ For the week-8 analysis, analyzing viral rebound in the first 8 weeks after ART treatment interruption, $C$ is the nadir CD4 count (categorized as $\leq 500$ versus $> 500$). Not available in 2 patients, who were excluded from the week-8 analysis.\\
$^d$ Difference in probability of virologic suppression that is mediated by binary mediator, expressed as a percentage.\\
$^e$ 95\% Confidence Interval, calculated by bootstrap with 5000 replicates using Efron's percentile method (Van~der~Vaart~(1998)\cite{Vaart} page~327). This leads to consistent coverage because of the bootstrap Z-estimator master theorem, Theorem~10.16 from Kosorok~(2008).\cite{Kosorokbook}\\
$^f$ Indirect effect effect on the Risk Ratio (RR) scale.\\
$^g$  Single-copy plasma HIV-RNA, on-ART. Analysis restricted to the 94 patients with SCA HIV-RNA measured.\\
$^h$ Cell-associated HIV-RNA, on-ART. All 124 patients had CA HIV-RNA measured.\\
$^i$ Treatment $A$ which causes all mediator values below the assay limit.\\
\end{table}

\begin{table}
\caption{Organic indirect effects of curative HIV-treatments that shift the distribution of the HIV-reservoir measures downwards.\\
The estimated probability of virologic suppression without treatment was $63/124$ or $51\%$ (week 4) and $17/122$ or $14\%$ (week 8). }
\begin{tabular}{lllllllll}
Mediator& Shift (log$_{10}$ scale)$^a$ & Week$^{b,c}$ & Indirect effect$^d$ & 95\% CI$^e$ & RR$^f$ & 95\% CI$^e$\\
\hline\\
SCA HIV-RNA$^g$ & 0.5 log$_{10}$ & 4 & 2.5\% & (-2.5\%,7.0\%) & 1.05 & (0.96,1,13)\\
SCA HIV-RNA$^g$ & 1 log$_{10}$ & 4 & 5.7\% & (-1.3\%,12.6\%)  & 1.10 & (0.98,1.24)\\
SCA HIV-RNA$^g$ & $\infty$$^i$ & 4 & 6.2\% & (-1.4\%,14.0\%) & 1.12 & (0.97,1.27)\\
\\
CA HIV-RNA$^h$& 0.5 log$_{10}$ & 4 & 6.9\% & (1.7\%,12.6\%) & 1.14 & (1.03,1.26)\\
CA HIV-RNA$^h$ & 1 log$_{10}$ & 4 & 9.8\% & (2.7\%,17.0\%) & 1.19 & (1.05,1.36)\\
CA HIV-RNA$^h$ & $\infty$$^i$ & 4 & 12.7\% & (3.0\%,22.7\%) & 1.25 & (1.06,1.47)\\
\\
SCA HIV-RNA$^g$ & 0.5 log$_{10}$ & 8 & 2.9\% & (0.61\%,6.3\%) & 1.18 & (1.04,1.40)\\
SCA HIV-RNA$^g$ & 1 log$_{10}$ & 8 & 5.1\% & (0.92\%,10.1\%)  & 1.32 & (1.07,1.62)\\
SCA HIV-RNA$^g$ & $\infty$$^i$ & 8 & 5.7\% & (0.95\%,11.1\%) & 1.35 & (1.07,1.68)\\
\\
CA HIV-RNA$^h$& 0.5 log$_{10}$ & 8 & 9.1\% & (3.5\%,15.5\%) & 1.66  & (1.29,2.15)\\
CA HIV-RNA$^h$ & 1 log$_{10}$ & 8 & 11.9\% & (4.9\%,19.4\%) & 1.85 & (1.40,2.45)\\
CA HIV-RNA$^h$ & $\infty$$^i$ & 8 & 15.9\% & (6.8\%,25.5\%) & 2.14 & (1.56,2.92)\\
\\
\end{tabular}\\
\label{tab:cont}
\noindent$^a$ Downwards shift of the mediator distribution, given the common cause $C$, on the log$_{10}$ scale, relative to no curative treatment $a=0$.\\
$^b$ For the week-4 analysis, analyzing viral rebound in the first 4 weeks after ART treatment interruption, $C$ is NNRTI-based (versus not).\\
$^c$ For the week-8 analysis, analyzing viral rebound in the first 8 weeks after ART treatment interruption, $C$ is the nadir CD4 count (categorized as $\leq 500$ versus $> 500$). Not available in 2 patients, who were excluded from the week-8 analysis.\\
$^d$ Difference in probability of virologic suppression that is mediated, expressed as a percentage.\\
$^e$ 95\% Confidence Interval, calculated by bootstrap with 5000 replicates using Efron's percentile method (Van~der~Vaart~(1998)\cite{Vaart} page~327). This leads to consistent coverage because of the bootstrap Z-estimator master theorem, Theorem~10.16 from Kosorok~(2008).\cite{Kosorokbook}\\
$^f$ Indirect effect effect on the Risk Ratio (RR) scale.\\
$^g$  Single-copy plasma HIV-RNA, on-ART. Analysis restricted to the 94 patients with SCA HIV-RNA measured.\\
$^h$ Cell-associated HIV-RNA, on-ART.  All 124 patients had CA HIV-RNA measured.\\
$^i$ Treatment $A$ which causes all mediator values below the assay limit.\\
\end{table}
\end{landscape}

\appendix

\section{The Mediation Formula for organic indirect and direct effects}\label{Amedform}

For completeness, in this eAppendix, we prove the Mediation Formula for organic indirect and direct effects relative to $a=0$. The proof follows along the same lines as the proof for organic indirect and direct effects relative to $a=1$ introduced in Lok (2016)\cite{medJL}, with the roles of $A=1$ and $A=0$ reversed.\\

\noindent {\bf Theorem}: \textbf{\emph{Organic indirect and direct effects: the Mediation Formula.}} Under randomized treatment, the following holds for an intervention $I$ that is organic relative to $a=0$ and $C$:
\begin{equation*}E\left(Y^{(0,I=1)}\right)
=\int_{(m,c)}E\left[Y|M=m,C=c,A=0\right]f_{M|C=c,A=1}(m)f_C(c)dm\,dc.
\end{equation*}

\noindent {\bf Proof:}
\begin{eqnarray*}
E\left(Y^{(0,I=1)}\right)
&=&E\left(E\left[Y^{(0,I=1)}|M^{(0,I=1)},C\right]\right)\\
&=&\int_{(m,c)}E\left[Y^{(0,I=1)}|M^{(0,I=1)}=m,C=c\right]f_{M^{(0,I=1)}|C=c}(m)dm\,f_C(c)dc\\
&=&\int_{(m,c)}E\left[Y^{(0)}|M^{(0)}=m,C=c\right]f_{M^{(1)}|C=c}(m)dm\,f_C(c)dc\\
&=&\int_{(m,c)}E\left[Y^{(0)}|M^{(0)}=m,C=c,A=0\right]f_{M^{(1)}|C=c,A=1}(m)dm\,f_C(c)dc\\
&=&\int_{(m,c)}E\left[Y|M=m,C=c,A=0\right]f_{M|C=c,A=1}(m)f_C(c)dm\,dc.
\end{eqnarray*}
The second equality follows because of the definition of conditional expectations. The third equality follows because of the two parts of the definition of organic interventions.
The fourth equality follows because under randomized treatment,
\begin{equation*}A\cip \bigl(Y^{(0)},M^{(0)}\bigr)\mid C \hspace{1cm}{\rm and}\hspace{1cm}A \mid \cip M^{(1)}|C.
\end{equation*}
The last equality follows because if an individual is randomized to $A=0$, we observe his/her outcome under $A=0$, and if an individual is randomized to $A=1$, we observe his/her outcome under $A=1$.

\section{Organic indirect and direct effects relative to $a=0$: the product method for linear models with treatment-mediator interaction: proof}\label{Aproduct}

In this eAppendix, we show that the product method\cite{BaronKenny} holds for organic indirect and direct effects relative to $a=0$ under linear models, regardless of whether there is an interaction between treatment and mediator in the outcome model.\\

\noindent {\bf Assumptions for the product method: linear models.}
\begin{equation*}M=\alpha_0+\alpha_1C+\alpha_2A+\epsilon_M,\end{equation*}
with $\epsilon_M\cip A\mid C$, and
\begin{equation*}Y=\beta_0+\beta_1C+\beta_2A+\beta_3M+\beta_4AM+\epsilon_Y,\end{equation*}
with $E[\epsilon_Y|M,A,C]=0$.\\

\noindent {\bf Theorem (product method for organic indirect effect relative to $a=0$).} Under the above linear models assumptions for the product method, the organic indirect effect relative to $a=0$ is equal to
\begin{equation*} E\bigl(Y^{(0,I=1)}\bigr)-E\bigl(Y^{(0)}\bigr)=\beta_3\alpha_2.
\end{equation*}

\noindent{\bf Proof:}
\begin{eqnarray*}E\bigl(Y^{(0,I=1)}\bigr)-E\bigl(Y^{(0)}\bigr)
&=&\int_{(m,c)}E\left[Y|M=m,A=0,C=c\right]\\
&&\;\;\left(f_{M|A=1,C=c}(m)-f_{M|A=0,C=c}(m)\right)f_C(c)dm\,dc\\
&=&\int_{(m,c)}(\beta_0+\beta_1c+\beta_3m)\left(f_{M|A=1,C=c}(m)-f_{M|A=0,C=c}(m)\right)f_C(c)\,dm\,dc\\
&=&\beta_3\int_{(m,c)}m\left(f_{M|A=1,C=c}(m)-f_{M|A=0,C=c}(m)\right)f_C(c)\,dm\,dc\\
&&+\int_c (\beta_0+\beta_1c) \bigl(\int_m \left(f_{M|A=1,C=c}(m)-f_{M|A=0,C=c}(m)\right)dm\bigr)\,dc\\
&=&\beta_3\int_{(m,c)}mf_{M|A=0,C=c}(m-\alpha_2)f_C(c)dm\,dc\\
&&-\beta_3\int_{(m,c)}mf_{M|A=0,C=c}(m)f_C(c)dm\,dc+0\\
&=&\beta_3\int_{(\tilde{m},c)}(\tilde{m}+\alpha_2)f_{M|A=0,C=c}(\tilde{m})f_C(c)d\tilde{m}\,dc\\
&&-\beta_3\int_{(m,c)}mf_{M|A=0,C=c}(m)f_C(c)dm\,dc\\
&=&\beta_3\alpha_2.
\end{eqnarray*}
The first equality follows from the Mediation Formula (first term) and a conditioning argument (second term). The fourth equality uses that $f_{M|A=1,C=c}(m)=f_{M|A=0,C=c}(m-\alpha_2)$. The fifth equality substitutes $\tilde{m}=m-\alpha_2$. That finishes the proof.

\section{Usefulness of combining intervention on the mediator with no treatment, $a=0$: mother-to-child transmission of HIV/AIDS}\label{AHIVtrans}

HIV-infected mothers can transmit the HIV-virus to their infants. The effect of AZT treatment on mother-to-child transmission of HIV-1 is surprisingly large, given the limited effect of AZT mono-therapy on HIV-1 RNA.\cite{mtctVictor} Less than $20\%$ of the effect of AZT on mother to child transmission can be explained through the effect of AZT on HIV-1 RNA.\cite{Sperling}

What is the likely effect on mother-to-child transmission of a potential new treatment that has the same effect on HIV-1 RNA as AZT but no direct effect on mother to child transmission? In this case, the outcome $Y$ is an indicator ``newborn baby is HIV-infected''. The mediator $M$ is HIV-1 RNA. $I$ is an intervention that, without AZT, causes the distribution of HIV-1
RNA, $M^{(0,I=1)}$, to be the same as under AZT; the potential new treatment. The quantity of interest is then $E\bigl(Y^{(0,I=1)}\bigr)-E\bigl(Y^{(0)}\bigr)$. This quantity is different from the usual indirect effect and similar to the pure indirect effect,\cite{RobGreenmed} see also VanderWeele~(2015)\cite{VanderWeelebook} pages~193-194: it combines an intervention $I$ on the mediator with no treatment, instead of with treatment.

The natural indirect and direct effect (or the pure indirect effect) are not very meaningful in this case, since the HIV-viral load in the mother's blood cannot be set. Once we will be able to set the HIV-viral load, we will set it to $0$. Thus, for a treatment like AZT, organic indirect and direct effects are more interpretable than their natural counterparts.

Combining organic interventions $I$ with no treatment provides information on what to expect from a treatment that affects the HIV-viral load in the mother's blood the same way as AZT does, but has no direct effect on mother-to-child transmission.

\section{Uniqueness of organic indirect and direct effects}\label{Aunique}

\noindent {\bf Definition:} (\emph{common cause}). $X$ is \emph{not} a common cause of mediator $M$ and outcome $Y^{(0)}$ given $C$ if either equation~(\ref{common1}) or equation~(\ref{common2}) holds:
\begin{equation}\label{common1}
X\cip M^{(0)}\mid C \hspace{2cm} {\rm and}\hspace{2cm} X\cip M^{(1)}\mid C
\end{equation}
or
\begin{equation} \label{common2}
X\cip Y^{(0)}\mid M^{(0)},C.
\end{equation}
In graphical language: $X$ is a common cause of mediator $M$ and outcome $Y^{(0)}$
if in a causal DAG that has $C$, $X$, $M$, and $Y^{(0)}$, there is an arrow from
$X$ to $M$, and there is a direct arrow from $X$ to $Y^{(0)}$. This
definition of common cause from Lok (2016)\cite{medJL} is in line with, for example, the concept of d-separation in Pearl (2000)\cite{Pea} pages~16-17.

In addition, we define:\\

\noindent {\bf Definition:} (\emph{all common causes}). A pre-treatment set of variables $C$ has all pre-treatment common causes of mediator $M$ and outcome $Y^{(0)}$ if all other pre-treatment sets of variables $X$ satisfy (\ref{common1}) and/or (\ref{common2}) with respect to $C$; that is, they are not a common cause of mediator $M$ and outcome $Y^{(0)}$ given $C$.\medskip

\noindent Such set $C$ is not unique.

Now, let $I^C$ be an intervention that is organic with respect to $C$ and let $I^{\tilde{C}}$ be an
intervention that is organic with respect to $\tilde{C}$. 

If both $C$ and $\tilde{C}$ have all common causes of mediator $M$ and outcome $Y^{(0)}$, then $C$ is not a common cause of mediator and outcome $Y^{(0)}$ given $\tilde{C}$, and $\tilde{C}$ is not a common cause of mediator $M$ and outcome $Y^{(0)}$ given $C$.
Hence there are 4 different cases, with either (\ref{common1}) or (\ref{common2}) holding for $C$ and $\tilde{C}$, respectively. Similar to Lok (2016),
\cite{medJL} it can be shown that under any of those 4 different cases,
\begin{equation*}
E\left(Y^{(0,I^{\tilde{C}}=1)}\right)
=\int_{(m,\tilde{c},c)}E\bigl[Y^{(0)}\mid M^{(0)}=m,\tilde{C}=\tilde{c},C=c\bigr]f_{M^{(1)}\mid \tilde{C}=\tilde{c},C=c}(m)f_{\tilde{C},C}(\tilde{c},c)dm\,d\tilde{c}\,dc.
\end{equation*}
Because of symmetry, it follows that also
\begin{equation*}
E\left(Y^{(0,I^{C}=1)}\right)
=\int_{(m,\tilde{c},c)}E\bigl[Y^{(0)}\mid M^{(0)}=m,\tilde{C}=\tilde{c},C=c\bigr]f_{M^{(1)}\mid \tilde{C}=\tilde{c},C=c}(m)f_{\tilde{C},C}(\tilde{c},c)dm\,d\tilde{c}\,dc.
\end{equation*}
But then, $E\left(Y^{(0,I^{\tilde{C}}=1)}\right)=E\left(Y^{(0,I^{C}=1)}\right)$.

\section{A weaker identifiability condition for natural indirect and direct effects}\label{Aweaker}

Consider settings where the mediator under treatment can be set to its value without treatment, and there is consensus about the closest possible world to this one where this can be accomplished. In such setting, the natural indirect and direct effects are well-defined. Now let $M^{(1,I=1)}=M^{(0)}$; this is an intervention $I$ on the mediator. If all $Y^{(a,m)}$ exist, this intervention $I$ is an organic intervention if:
\begin{equation*}
Y^{(1,I=1)}\mid M^{(1,I=1)}=m,C=c\;\;\;\;\;\sim\;\;\;\;\;Y^{(1)}\mid M^{(1)}=m,C=c,
\end{equation*}
or equivalently, since in this particular example $M^{(1,I=1)}=M^{(0)}$ and all $Y^{(a,m)}$ are well-defined,
\begin{equation}\label{natcond}
Y^{(1,m)}\mid M^{(0)}=m,C=c\;\;\;\;\;\sim\;\;\;\;\;Y^{(1,m)}\mid M^{(1)}=m,C=c.
\end{equation}
Analogous to Lok (2016)\cite{medJL} Theorem~4.4, this follows e.g., under the usual conditions for identification of natural indirect and direct effects. Then, the natural indirect and direct effects are the same as the organic indirect and direct effects.

Still, equation~(\ref{natcond}) is a cross-worlds assumption, with $Y^{(1,m)}$ conditional on $M^{(0)}$. This is no surprise, since the definition of natural indirect and direct effects relies on cross-worlds quantities. The above does relax the usual assumptions for identification of natural indirect and direct effects.

\section{Indirect effects relative to $a=0$ estimated from the treatment effect on the mediator and outcome data without treatment}\label{Aexa}

For the situation where both mediator $M$ and outcome $Y$ follow linear models, we have the product method for the indirect effect relative to $a=0$, see eAppendix~\ref{Aproduct}.

In general, in equation~(\ref{indir}), we derived that the organic indirect effect of a treatment relative to $a=0$ is equal to
\begin{equation*}\int_{(m,c)}E\left[Y|M=m,C=c,A=0\right]\left(f_{M|C=c,A=1}(m)-f_{M|C=c,A=0}(m)\right)f_C(c)dm\,dc.
\end{equation*}
In this eAppendix, we present some examples on how to use this formula to estimate the organic indirect effect relative to $a=0$. 

First, for binary mediators $M$, we prove that the organic indirect effect relative to $a=0$ equals
\begin{eqnarray*}\lefteqn{\int_{c}\left(E\left[Y|M=1,C=c,A=0\right]-E\left[Y|M=0,C=c,A=0\right]\right)}\nonumber\\
&&\left(P(M=1|C=c,A=1)-P(M=1|C=c,A=0)\right)f_C(c)\,dc
\end{eqnarray*}
(see equation~(\ref{productBin})); a product method for binary mediators. We derive this starting with the above equation~(\ref{indir}):
\begin{eqnarray*}\lefteqn{\int_{(m,c)}E\left[Y|M=m,C=c,A=0\right]\left(f_{M|C=c,A=1}(m)-f_{M|C=c,A=0}(m)\right)f_C(c)dm\,dc}\\
&=&\int_{c}E\left[Y|M=1,C=c,A=0\right]\left(P(M=1|C=c,A=1)-P(M=1|C=c,A=0)\right)f_C(c)\,dc\\
&&+\int_{c}E\left[Y|M=0,C=c,A=0\right]\left(P(M=0|C=c,A=1)-P(M=0|C=c,A=0)\right)f_C(c)\,dc\\
&=&\int_{c}E\left[Y|M=1,C=c,A=0\right]\left(P(M=1|C=c,A=1)-P(M=1|C=c,A=0)\right)f_C(c)\,dc\\
&&-\int_{c}E\left[Y|M=0,C=c,A=0\right]\left(P(M=1|C=c,A=1)-P(M=1|C=c,A=0)\right)f_C(c)\,dc\\
&=&\int_{c}\left(E\left[Y|M=1,C=c,A=0\right]-E\left[Y|M=0,C=c,A=0\right]\right)\\
&&\;\hspace{2cm}\left(P(M=1|C=c,A=1)-P(M=1|C=c,A=0)\right)f_C(c)\,dc.
\end{eqnarray*}
With a similar derivation, the organic indirect effect relative to $a=1$ equals
\begin{eqnarray*}\lefteqn{\int_{c}\left(E\left[Y|M=1,C=c,A=1\right]-E\left[Y|M=0,C=c,A=1\right]\right)}\nonumber\\
&&\left(P(M=1|C=c,A=1)-P(M=1|C=c,A=0)\right)f_C(c)\,dc;
\end{eqnarray*}
a product method for binary mediators $M$ also for what is usually called the natural indirect effect.

As another example,  suppose that $Y$ without treatment is continuously distributed, and follows a regression model under no treatment such as
\begin{equation*}Y=\beta_0+\beta_1C+\beta_2M+\beta_3CM+\epsilon_Y,\end{equation*}
with $E[\epsilon_Y|M,C]=0$. Or, $Y$ under no treatment could follow a logistic regression model such as
\begin{equation*} \rm{logit} \;P(Y=1|M=m,C=c,A=0)=\beta_0+\beta_1C+\beta_2 M +\beta_3 CM.\end{equation*}
More generally, $Y$ could be continuous, binary, or e.g., counts, and follow a generalized linear model such as
\begin{equation}E\left[Y|M=m,C=c,A=0\right]=g^{-1}(\beta_1C+\beta_2M+\beta_3CM),\label{glm}\end{equation}
with $g$ any link function. This includes both regression and logistic regression, so we will work with (\ref{glm}) below. 

An estimator for the organic indirect effect relative to $a=0$ that can always be used in randomized studies if equation~(\ref{glm}) holds is
\begin{eqnarray}\label{general}\lefteqn{\frac{1}{\# {\rm treated}}\sum_{i:\, {\rm person }\;i\;{\rm treated}}^n g^{-1}(\hat{\beta}_1c_i+\hat{\beta}_2m_i+\hat{\beta}_3c_im_i)}\nonumber\\
&&-\frac{1}{\# {\rm untreated}}\sum_{i:\, {\rm person }\;i\;{\rm untreated}}^n g^{-1}(\hat{\beta}_1c_i+\hat{\beta}_2m_i+\hat{\beta}_3c_im_i).
\end{eqnarray}
The ingredients in this expression can easily be estimated using most statistical software packages. First, (\ref{glm}) is fitted using common causes, mediator, and outcome data under no treatment ($A_i=0$), resulting in $\hat{\beta}$. Then the mediator data under both no treatment and under treatment can be plugged into (\ref{general}).

Next, we present two specific examples. As a first specific example, if (\ref{glm}) holds under no treatment and the mediator $M$ is binary, using (\ref{productBin}), the organic indirect effect relative to $a=0$ of a treatment is equal to
\begin{equation*}
\int_{c}\left(g^{-1}(\beta_1c+\beta_2+\beta_3c)-g^{-1}(\beta_1c)\right) (P(M=1|C=c,A=1)-P(M=1|C=c,A=0))f_C(c)\,dc,
\end{equation*}
which under the usual regularity conditions can be consistently estimated by
\begin{equation*}\frac{1}{n}\sum_{i=1}^n\left(g^{-1}(\hat{\beta}_1c_i+\hat{\beta}_2+\hat{\beta}_3c_i)-g^{-1}(\hat{\beta}_1c_i) \right) (\hat{P}(M=1|C=c_i,A=1)-\hat{P}(M=1|C=c_i,A=0)).
\end{equation*}
Here, $\hat{P}(M=1|C=c_i,A=1)$ and $\hat{P}(M=1|C=c_i,A=0)$ can be estimated using one's favorite model for binary variables.

As a second specific example, consider settings where (\ref{glm}) holds under no treatment and the mediator $M$ is continuous and follows a regression model
\begin{equation*}M=\alpha_0+\alpha_1C+\alpha_2A+\epsilon_M,\end{equation*}
with $\epsilon_M\cip A\mid C$. 
This mediator model could e.g.\ be estimated from mediator data under both no treatment and under treatment. In such setting, the organic indirect effect relative to $a=0$ of treatment $A$ is equal to
\begin{eqnarray*}\lefteqn{\int_{(m,c)} g^{-1}(\beta_1c+\beta_2m+\beta_3cm)\left(f_{M|C=c,A=0}(m-\alpha_2)-f_{M|C=c,A=0}(m)\right)f_C(c)dm\,dc}\\
&=&\int_{(m,c)} g^{-1}(\beta_1c+(\beta_2+\beta_3c)(\tilde{m}+\alpha_2))f_{M|C=c,A=0}(\tilde{m})f_C(c)dm\,dc
\\
&&-\int_{(m,c)} g^{-1}(\beta_1c+(\beta_2+\beta_3c)m)f_{M|C=c,A=0}(m)f_C(c)dm\,dc\\
&=&\int_{(m,c)} g^{-1}(\beta_1c+(\beta_2+\beta_3c)(m+\alpha_2))f_{M|C=c,A=0}(m)f_C(c)dm\,dc
\\
&&-\int_{(m,c)} g^{-1}(\beta_1c+(\beta_2+\beta_3c)m)f_{M|C=c,A=0}(m)f_C(c)dm\,dc,
\end{eqnarray*}
where we used that $f_{M|A=1,C=c}(m)=f_{M|A=0,C=c}(m-\alpha_2)$ and we introduced $\tilde{m}=m-\alpha_2$. Under the usual regularity conditions this can be consistently estimated by e.g.:
\begin{equation*}\frac{\sum_{i: \,{\rm person }\;i\;{\rm untreated}}^n g^{-1}(\hat{\beta}_1c_i+(\hat{\beta}_2+\hat{\beta}_3c_i)(m_i+\hat{\alpha}_2))-g^{-1}(\hat{\beta}_1c_i+(\hat{\beta}_2+\hat{\beta}_3c_i)m_i)}{\# {\rm untreated}}.
\end{equation*}

In our HIV-application, we consider a new treatment which, given the pre-treatment common causes $C$, shifts the distribution of the log$_{10}$ HIV-persistence measure to the left by one-log$_{10}$, or more generally, $\alpha$ log$_{10}$. For such shift, assuming that the generalized linear model (\ref{glm}) holds for the outcome $Y$, the indirect effect relative to $a=0$ of the new treatment mediated by the HIV-persistence measure $M$ equals
\begin{eqnarray*}\lefteqn{\int_{(m,c)} g^{-1}(\beta_1c+\beta_2m+\beta_3cm)\left(f_{M|C=c,A=0}(m+\alpha)-f_{M|C=c,A=0}(m)\right)f_C(c)dm\,dc}\\
&=&\int_{(m,c)} g^{-1}(\beta_1c+(\beta_2+\beta_3c)(\tilde{m}-\alpha))f_{M|C=c,A=0}(\tilde{m})f_C(c)dm\,dc
\\
&&-\int_{(m,c)} g^{-1}(\beta_1c+(\beta_2+\beta_3c)m)f_{M|C=c,A=0}(m)f_C(c)dm\,dc\\
&=&\int_{(m,c)} g^{-1}(\beta_1c+(\beta_2+\beta_3c)(m-\alpha))f_{M|C=c,A=0}(m)f_C(c)dm\,dc
\\
&&-\int_{(m,c)} g^{-1}(\beta_1c+(\beta_2+\beta_3c)m)f_{M|C=c,A=0}(m)f_C(c)dm\,dc.
\end{eqnarray*}
We can estimate the organic indirect effect relative to $a=0$ of the new treatment mediated by the log HIV-persistence measure in 3 steps.  1.\ Assume a particular value of $\alpha$, or estimate $\alpha$ for a particular new treatment. 2.\ Use mediator and outcome data under no treatment to fit (\ref{glm}) and obtain $\hat{\beta}$. 3.\ Estimate the organic indirect effect relative to $a=0$ of the new treatment mediated by the log HIV-persistence measure by
\begin{equation*} \frac{\sum_{i: \,{\rm person }\;i\;{\rm untreated}}^n g^{-1}(\hat{\beta}_1c_i+(\hat{\beta}_2+\hat{\beta}_3c_i)(m_i-\hat{\alpha})) - g^{-1}(\hat{\beta}_1c_i+(\hat{\beta}_2+\hat{\beta}_3c_i)m_i)}{\# {\rm untreated}}.
\end{equation*}

\section{HIV application: what if we have on-treatment measurements of the mediator $M$ in a different populaton?}\label{Mmeasurements}

If the pre-treatment characteristics are different in the two populations, matching (e.g., Rosenbaum (2017)\cite{Rosenbaumbook} Chapter~11) could be used on top, or inverse probability of treatment weighting.\cite{MSM1,MSM2} Alternatively, one could choose a distribution $f_C(c)$ of $C$ of interest, fit a model for the distribution of $M^{(0)}$ given $C$ (e.g., eTable~\ref{tab:M}) and $M^{(1)}$ given $C$ (from $(m^{(1)}_{j},c^{(1)}_{j})_{j=1}^{J}$), and estimate the organic indirect effect in this population by
\begin{equation*}\int_{(m,c)}p_{\hat{\beta}}(m,c)\left(\hat{f}_{M|C=c,A=1}(m)-\hat{f}_{M|C=c,A=0}(m)\right)f_C(c)\,dm\,dc.
\end{equation*}
If the distribution $f_C(c)$ of $C$ of interest is that of our study population, the organic indirect effect can be estimated by
\begin{equation*}\frac{1}{n}\sum_{i=1}^n \left(\int_{m} p_{\hat{\beta}}(m,c_i)\hat{f}_{M|C=c_i,A=1}(m)\,dm-y_i\right).
\end{equation*}

\section{Multiplying the odds}\label{Odds}

In this Web-appedix, we show that if treatment increases the odds of a binary variable $\bar{M}$ being $1$ by a factor F, then if $P(\bar{M}^{(0)}=1)=p_0$ without treatment, $P(\bar{M}^{(1)}=1)=Fp_0/(1-p_0+Fp_0)$ under treatment. For $F=2$, this implies that $P(\bar{M}^{(1)}=1)=2p_0/(1+p_0)$.

Write $p_1=P(\bar{M}^{(1)}=1)$. Then if the odds of $1$ increases by a factor $F$ due to treatment,
\begin{eqnarray*}
\frac{p_1}{1-p_1}=F\frac{p_0}{1-p_0}
&\Rightarrow & p_1=\frac{F(1-p_1)p_0}{1-p_0}\\
&\Rightarrow & p_1\left(1+F\frac{p_0}{1-p_0}\right)=F\frac{p_0}{1-p_0}\\
&\Rightarrow & p_1\frac{1-p_0+Fp_0}{1-p_0}=\frac{Fp_0}{1-p_0}\\
&\Rightarrow & p_1(1-p_0+Fp_0)=Fp_0\\
&\Rightarrow & p_1=\frac{Fp_0}{1-p_0+Fp_0}.
\end{eqnarray*}

\section{Observational data}\label{Aobs}

With observational data, the definition of organic indirect and direct effects should not change. As in Lok (2016),\cite{medJL} an identification result holds for observational data similar to the identification result for randomized data, provided that $C$ has all common causes of mediator and outcome.

There may exist baseline covariates $Z$ (beyond the common causes $C$ of mediator and outcome)
that need to be included in the analysis in order to eliminate confounding:\\

\noindent {\bf Assumption:} \emph{(No Unmeasured Confounding).}
\begin{equation*}A\cip \bigl(Y^{(0)},M^{(0)}\bigr)\mid C,Z \;\;\;\;\;{\rm and}\;\;\;\;\; A\cip \bigl(Y^{(1)},M^{(1)}\bigr)\mid C,Z.
\end{equation*}

\noindent We adopt the usual Consistency Assumption relating the observed data to the counterfactual data:\\

\noindent {\bf Assumption:} \emph{(Consistency).} On $A=1$, $M=M^{(1)}$ and $Y=Y^{(1)}$. On $A=0$, $M=M^{(0)}$ and $Y=Y^{(0)}$.\\

\noindent {\bf Theorem:} \emph{(Organic indirect and direct effects: the
  Mediation Formula for observational data).} Assume No Unmeasured
Confounding, Consistency, intervention $I$ is organic with respect to $C$, and given $C$, $Z$ is not a common cause of mediator and outcome (see eAppendix~\ref{Aunique}). Then
\begin{equation*}
E\bigl(Y^{(0,I=1)}\bigr)
=\int_{(m,c,z)}E\left[Y|M=m,C=c,Z=z,A=0\right]f_{M|C=c,Z=z,A=1}(m)f_{C,Z}(c,z)dm\,d(c,z).
\end{equation*}
A similar Mediation Formula for observational data holds for $E\left(Y^{(1,I=1)}\right)$, by reversing the roles of $a=0$ and $a=1$. The proof of this Mediation Formula is similar to the proofs in eAppendix~\ref{Amedform} and in Lok (2016).\cite{medJL} 

\section{HIV-application: Models fit to estimate the indirect effects}\label{AppHIVResults}

This eAppendix describes the models fit to estimate the indirect effects relative to $a=0$ in Section~\ref{HIV}. Viral rebound was defined as the time of the first viral load measurement above $1000$ copies/ml after ART interruption. %above is good, I programmed >1000
In the week-4 analysis, with outcome viral rebound within 4 weeks, we included NNRTI use as a common cause $C$ of mediator $M$ and outcome $Y$. NNRTI disappears from the system after 8 weeks.\cite{Li2015} %Their page 8. Ron says also in Julg but couldn't get Julg fast so decided this is fine.
In the week-8 analysis, we included the nadir CD4 count as a common cause $C$ of mediator $M$ and outcome $Y$.

For the mediator cell-associated HIV-RNA (CA HIV-RNA), the assay limit was 92 copies /million CD4+ T-cells. For the mediator single-copy plasma HIV-RNA (SCA HIV-RNA), since it was a single copy assay, we took the assay limit to be $1$ copies/ml, consistent with Figures~3e and~3f in Li et al (2016).\cite{Li2015} %checked and this is indeed what we did.

\begin{table}
\caption{Odds Ratio Estimates of virologic suppression ($Y=1$) given the mediator and pre-treatment common causes}
\begin{tabular}{llll}
week & predictor &OR$^a$& 95\% Wald CI$^b$\\
\hline\\
4 & Pre-ATI SCA HIV-RNA$^c$ below$^e$ & 2.1 & (0.86,5.0)\\ 
&NNRTI-based ART$^e$ & 2.3 & (1.003,5.5)\\ 
\\
4 & Pre-ATI SCA HIV-RNA$^c$ below$^e$ & 2.3 & (0.58,9.4)\\ 
& Continuous Pre-ATI SCA HIV-RNA$^c$ & 1.2 & (0.20,7.4)\\
& NNRTI-based ART$^e$ & 2.4 & (1.007,5.6)\\ 
\\
4 & Pre-ATI CA HIV-RNA$^d$ below$^e$ & 2.7 & (1.2,6.0)\\
& NNRTI-based$^e$ & 3.9 & (1.8,8.6)\\
\\
4 & Pre-ATI CA HIV-RNA$^d$ below$^e$ & 0.67 & (0.048,9.3)\\ 
& Continuous Pre-ATI CA HIV-RNA$^d$ & 0.58 & (0.22,1.54)\\
& NNRTI-based$^e$ & 4.0 & (1.8,8.7)\\ 
\\
8 & Pre-ATI SCA HIV-RNA$^c$ below$^e$ & 4.6 & (0.945,22.2)\\
& Nadir CD4 $\leq$ 500$^e$ & 4.6 & (0.945,22.2)\\
\\
8 & Pre-ATI SCA HIV-RNA$^c$ below$^e$ & 4.1 & (0.30,57)\\ 
& Continuous Pre-ATI SCA HIV-RNA$^c$ & 0.83 & (0.015,45)\\
& Nadir CD4 $\leq$ 500$^e$ &  4.5 & (0.93,22.2)\\ 
\\
8 & Pre-ATI CA HIV-RNA$^d$ below$^e$ & 9.0 & (2.5,32)\\
& Nadir CD4 $\leq$ 500$^e$ & 6.0 & (1.5,24)\\ 
\\
8 & Pre-ATI CA HIV-RNA$^d$ below$^e$ & 0.36 & ($<$0.001,208)\\ 
& Continuous Pre-ATI CA HIV-RNA$^d$ & 0.27 & (0.019,3.8)\\
& Nadir CD4 $\leq$ 500$^e$ & 5.9 & (1.45,24)\\ 
\\
\end{tabular}\\
$^a$ Odds Ratio.\\
$^b$ 95\% Wald Confidence Interval.\\
$^c$  Single-copy plasma HIV-RNA, on ART. Analysis restricted to the 94 patients with SCA HIV-RNA measured.\\
$^d$ Cell-associated HIV-RNA, on ART. All 124 patients had CA HIV-RNA measured.\\
$^e$: yes versus no.
\end{table}

\begin{table}
\caption{Odds Ratio Estimates of the mediator below the assay limit given the pre-treatment common cause}
\begin{tabular}{llll}
Mediator &Common cause &OR$^a$& 95\% Wald CI$^b$\\ 
\hline\\
Pre-ATI SCA HIV-RNA$^c$ below$^{e}$ & NNRTI-based ART$^e$ & 1.9 & (0.81,4.5)\\
Pre-ATI CA HIV-RNA$^d$ below$^{e}$ & NNRTI-based ART$^e$ & 0.65 & (0.32,1.3)\\ 
Pre-ATI SCA HIV-RNA$^c$ below$^{e}$ & Nadir CD4 $\leq$ 500$^e$ & 1.03 & (0.43,2.5)\\
Pre-ATI CA HIV-RNA$^d$ below$^{e}$ & Nadir CD4 $\leq$ 500$^e$ & 0.30 & (0.14,0.64)\\
\\ 
\end{tabular}\\
\label{tab:M}
$^a$ Odds Ratio.\\
$^b$ 95\% Wald Confidence Interval.\\
$^c$  Single-copy plasma HIV-RNA, on ART. Analysis restricted to the 94 patients with SCA HIV-RNA measured.
Without treatment, the probability of Pre-ATI SCA HIV-RNA below the assay limit was 60/94, or 64\%.\\
$^d$ Cell-associated HIV-RNA, on ART. All 124 patients had CA HIV-RNA measured.
Without treatment, the probability of Pre-ATI CA HIV-RNA below the assay limit was 52/124, or 42\%.\\
$^e$: yes versus no.
\end{table}

\newpage

For the week 8 analysis, many of the bootstrap samples had an outcome model with almost complete separation of data points, so the confidence intervals for the week 8 analyses might not be as reliable as those for the week 4 analysis. This is likely due to the limited number of patients with virologic suppression at week 8. For our analyses, we don’t need the parameter estimates in the bootstrap samples, only the probabilities for the outcome given the predictors, so an almost complete separation of data points might not be as problematic as in cases where the focus is on parameter estimates. The histograms of the bootstrap indirect estimates look like proper bell curves for all our analyses, although some of them are a bit skewed to the right.

\end{document}